\documentclass[preprint,NumberedRefs]{JASAnew}
\usepackage{siunitx} 
\usepackage{blindtext}
\usepackage{float}
\usepackage{upgreek}
 \usepackage{graphicx}
 \usepackage{booktabs}
\begin{document}

\title[\textit{JASA}/Cavitation in Bovine Liver]{An Ultrasonically Actuated Fine-Needle Creates Cavitation in Bovine Liver}

\author{Emanuele Perra}
\email{emanuele.perra@aalto.fi}
\thanks{Corresponding Author}
\affiliation{Medical Ultrasonics Laboratory (MEDUSA), Department of Neuroscience and Biomedical Engineering, Aalto University, Espoo, 02150, Finland}{}

\author{Nick Hayward}
\affiliation{Medical Ultrasonics Laboratory (MEDUSA), Department of Neuroscience and Biomedical Engineering, Aalto University, Espoo, 02150, Finland}{}

\author{Kenneth P.H. Pritzker}
\affiliation{Department of Laboratory Medicine and Pathobiology, University of Toronto, Toronto, M5S 1A8, Canada}
\affiliation{Department of Pathology and Laboratory Medicine, Mount Sinai Hospital, Toronto, M5G 1X5, Canada}

\author{Heikki J. Nieminen}
\affiliation{Medical Ultrasonics Laboratory (MEDUSA), Department of Neuroscience and Biomedical Engineering, Aalto University, Espoo, 02150, Finland}{}

\preprint{Emanuele Perra, \textit{JASA}}		

\date{\today} 

\begin{abstract}
Ultrasonic cavitation is being used in medical applications as a way to influence matter, such as tissue or drug vehicles, on a micro-scale. Oscillating or collapsing cavitation bubbles provide transient mechanical force fields, which can, \textit{e.g}., fractionate soft tissue or even disintegrate solid objects such as calculi. Our recent study demonstrates that an ultrasonically actuated medical needle can create cavitation phenomena inside water. However, the presence and behavior of cavitation and related bioeffects in diagnostic and therapeutic applications with ultrasonically actuated needles are not known. Using simulations, we demonstrate numerically and experimentally the cavitation phenomena near ultrasonically actuated needles. We define the cavitation onset within a liver tissue model with different total acoustic power levels. We directly visualize and quantitatively characterize cavitation events generated by the ultrasonic needle in thin fresh bovine liver sections enabled by high speed imaging. On a qualitative basis, the numerical and experimental results show a close resemblance in threshold and spatial distribution of cavitation. These findings are crucial for developing new methods and technologies employing ultrasonically actuated fine-needles such as ultrasound-enhanced fine-needle biopsy, drug delivery and histotripsy.
\end{abstract}

\maketitle

\section{Introduction}
In recent years, ultrasonic cavitation has emerged in various medical applications as a way to influence matter non-invasively. Cavitation is a phenomenon that can be described as the interaction between small spherical gas bubbles and pressure perturbations taking place in a medium. When the peak rarefactional pressure amplitude (PRPA) of an ultrasound field is low enough, gas bubbles can undergo stable oscillations about their equilibrium radius, which is usually referred to as stable cavitation\cite{Neppiras1980}. However, at elevated PRPAs, if certain threshold conditions are met\cite{Holland1990}, gas bubbles can collapse giving rise to transient cavitation\cite{Apfel1981}. The collapse of a cavitation bubble may generate different nonlinear acoustic phenomena in the surrounding medium, such as generation of rapid liquid microjets, acoustic emission in the form of shock waves and formation of high stress fields. These physical effects have been widely investigated and employed in different medical applications with the intent to, \textit{e.g.}, ablate tumors\cite{Kennedy2005}, fractionate calculi\cite{Matsumoto2005} or tissue\cite{Maxwell2009a} and enhance the permeability of cells for drug delivery applications\cite{Lentacker2014}. 

In our recent study, we have demonstrated that cavitation events can be generated in water by an ultrasonically actuated medical needle \cite{Perra2021}. Moreover, it has been shown that at $\sim$\SI{30}{\kHz}, ultrasound-enhanced fine-needle aspiration biopsy (USeFNAB) enhances the yield of a biopsy by 3-5$\times$ in liver compared to when a fine-needle aspiration biopsy (FNAB) procedure is conducted using a similar needle. These results suggested that the nonlinear acoustic phenomena generated at the needle tip, including cavitation, might play an important role in the tissue cutting mechanism in the context of biopsy applications and beyond. However, the potential presence of cavitation and related bioeffects in diagnostic and therapeutic applications with ultrasonically actuated needles require a more thorough understanding.

Actuation of medical needles by ultrasound is not a new concept, and a number of studies related to the topic can be found in the literature\cite{Gui2014a,Kuang2016,Sadiq2014a,Liao2014,Liao2013}. However, the applications have been limited to improve the needle visibility in ultrasound-guided regional anaesthesia and tissue biopsy\cite{Gui2014a,Kuang2016,Sadiq2014a} or to reduce the penetration force of a standard needle\cite{Liao2014,Liao2013}. So far, no research seems to have been conducted on studying the potential generation of nonlinear acoustic phenomena and their interaction with soft tissue.

In this study, we aim at studying the influence of cavitation on soft tissue under the action of an ultrasonically actuated needle. Numerical modeling is first used to simulate the time-dependent acoustic field generated by the ultrasonic needle and the cavitation bubble dynamics in a liver tissue model. The numerical results provided fundamental understanding on the cavitation nucleation threshold, spatial distribution and maximum size of the cavitation bubbles, and their influence on the tissue, according to a cavitation/tissue interaction model proposed by Mancia \textit{et al.}\cite{Mancia2019}. Experimentally, we developed a method to visualize cavitation bubbles in thin portions of fresh liver tissue, involving high-speed (HS) imaging using light transmission. Such understanding is crucial for optimising the safety and efficacy of clinical interventional procedures, including many for diagnostics and cancer treatments\cite{Blana2004a}.

\section{Methods}

\subsection{Numerical Simulation}
The computational software COMSOL Multiphysics v5.5 \cite{Multiphysics2020} was used to solve the different equations governing the cavitation bubble dynamics taking place in soft tissue. We assumed a scenario where a 21G $\times$ \SI{80}{mm} hypodermic needle is used, since it represents a common medical needle normally employed in FNAB applications. The needle is partially placed into a \SI{10}{mm} $\times$ \SI{12}{mm} cylinder representing a liver tissue sample and actuated at the ultrasonic frequency of \SI{33}{\kHz} and total acoustic power (TAP) of 0.2, 0.5 and \SI{0.8}{W}. Ultrasonic flexural standing waves are enabled in the needle shaft via an ultrasonic device with similar geometry and material properties to the one employed in the actual experiments (Fig. \ref{figure1}). 

The displacement field in the needle and the acoustic pressure field inside the tissue domain are first calculated in the time domain (time step = $\frac{1}{100f} = \SI{0.3}{\micro\second}$, simulated time = $2/f$ = \SI{60.60}{\micro\second}). In order to study the bubble dynamics, the Keller–Miksis equation\cite{Keller1980} (Eq. (\ref{kellermiksis})) is then solved in the liver domain as a global ODE, by having the surroundings of the needle seeded with cavitation nuclei periodically spaced by \SI{50}{\micro\meter}. The initial size $R_0$ of the nuclei was assumed to be \SI{500}{\nano\meter}, according to previous studies on cavitation behaviour in soft tissue\cite{Vlaisavljevich2015}. The acoustic forcing pressure $p_f$ is given by the pressure field calculated in a separate study step, hence, the bubble motion is assumed not to contribute to the total pressure field. For simplicity, the inter-bubble interaction is neglected.

The three-dimensional model was meshed with free tetrahedral elements, considering at least 20 nodes per wavelength, which was considered an appropriate number for minimizing the local approximation errors \cite{Thompson2006}. A detailed list of the model parameters is given in Table \ref{Table1}. 

\subsubsection{Acoustic Wave Propagation in Tissue}
The acoustic wave propagation in soft tissue has been modeled by adopting the Westervelt equation\cite{hamilton1998nonlinear}:
\begin{equation}
\begin{aligned}
&\nabla^2 p -\frac{1}{c_{\infty}^2}\frac{\partial^2p}{\partial t^2} + \frac{\delta}{c_{\infty}^4}\frac{\partial^3p}{\partial t^3}+\frac{\beta}{\rho_{\infty}c_{\infty}^4}\frac{\partial^2p}{\partial t^2}= 0,
\end{aligned}
\label{westervelt}
\end{equation}
where $p$ is the pressure, $c_{\infty}$ and $\rho_{\infty}$ are the speed of sound and the density of the medium, respectively. The first two terms of Eq. (\ref{westervelt}) describe the linear lossless propagation of sound in a medium, while the third term is associated with viscous losses, and the last one accounts for the nonlinear propagation in soft tissue. The sound diffusivity $\delta$ is defined as \cite{Solovchuk2013,Demi2014}:
\begin{equation}
\begin{aligned}
&\delta = \frac{2c_{\infty}^3\alpha}{\omega^2},
\end{aligned}
\label{diffusivity}
\end{equation}
where $\omega$ is the angular frequency, $\alpha$ is the acoustic absorption coefficient. The term $\beta$ represents the nonlinearity coefficient, expressed as:
\begin{equation}
\begin{aligned}
&\beta = 1+\frac{B}{2A},
\end{aligned}
\label{beta}
\end{equation}
with $B/A$ being the nonlinearity parameter given for a specific material. 

\subsubsection{Cavitation Model}
The Keller–Miksis equation\cite{Keller1980} was used to describe the bubble dynamics in soft tissue: 

\begin{equation}
\begin{aligned}
\left(1 - \frac{\dot{R}}{c_{\infty}}\right)R\ddot{R} +\frac32 \left(1 -\frac{\dot{R}}{3c_{\infty}}\right)\dot{R}^2 = \frac{1}{\rho_{\infty}}\Biggl(1+\frac{\dot{R}}{c_{\infty}}+
\frac{R}{c_{\infty}}\frac{d}{dt}\Biggr)\times\left[p_B-(p_{\infty}+p_f(t))-\frac{2S}{R}+J\right].
\end{aligned}
\label{kellermiksis}
\end{equation}
In the above, $R$, $\dot{R}$ and $\ddot{R}$ denote the radial displacement, velocity and acceleration of the cavitation bubble wall, respectively, and the constants $c_\infty$ and $\rho_\infty$ denote the speed of sound and the density of the medium. The driving pressure is expressed by $p_{f}(t)$, while the pressure at the air-liquid interface of the bubble is defined as\cite{Gaitan2005}:
\begin{equation}
\begin{aligned}
p_B = p_0 \left(\frac{R_0}{R}\right)^{3\kappa},
\end{aligned}
\label{pB}
\end{equation}
where $R_0$ the bubble radius at rest and $\kappa$ is the polytropic exponent. The term $p_0$ represents the internal pressure of the bubble when the bubble is at equilibrium, expressed as: 
\begin{equation}
\begin{aligned}
 p_0 = p_{\infty} +2\frac{S}{R_0},
\end{aligned}
\label{p0}
\end{equation}
where $p_{\infty}$ indicates the ambient pressure and $S$ the surface tension of the bubble. Eq. (\ref{kellermiksis}) is combined with the Kelvin–Voigt model\cite{Gaudron2015}, which leads the integral of the deviatoric stress $J$, accounting for the viscoelastic behaviour of soft tissue, to be expressed as follows\cite{Mancia2017}:

\begin{equation}
\begin{aligned}
J= 2\int_{R}^{\infty}{\frac{\tau_{rr}-\tau_{\theta\theta}}{r}dr} = \frac{-4\mu\dot{R}}{R}-\frac G2 \left[5-4\left(\frac{R_0}{R}\right)-\left(\frac{R_0}{R}\right)^4\right],
\end{aligned}
\label{J}
\end{equation}
where $\mu$ is the tissue viscosity and $G$ is the tissue shear modulus. $\tau_{rr}$ and $\tau_{\theta\theta}$ represent the the radial and tangential stresses, respectively, due to the bubble deformation. They are related as follows:
\begin{equation}
\begin{aligned}
\tau_{rr} = -4\mu\frac{R^2\dot{R}}{r^3}+2G\left[ \Biggl(\frac{r_0}{r}\Biggr)^4 -\Biggl(\frac{r}{r_0}\Biggr)^2\right] = -2\tau_{\theta\theta},
\end{aligned}
\label{trr}
\end{equation}
where $r$ is the radial coordinate and \(r_0 = \sqrt[3]{r^3 - R^3 - R_0^3}\) relates the coordinate $r$ to its initial position in the undeformed configuration of the surrounding tissue. 
The strain field in the surrounding tissue is defined as\cite{Mancia2019,Estrada2017}:
\begin{equation}
\begin{aligned}
E_{rr} = -2\text{ln}\left(\frac{r}{r_0}\right) = -2E_{\theta\theta},
\end{aligned}
\label{Err}
\end{equation}
being $E_{rr}$ and $E_{\theta\theta}$ the radial and tangential strain, respectively.

The model for cavitation/tissue interaction proposed by Mancia \textit{et al.} (2019) was adopted to estimate the amount of tissue volume influenced by the cavitation activity. Specifically, this is evaluated by identifying the regions where the von Mises strain (Eq. (\ref{Emis})) exceeds the ultimate fractional strain measured for liver (\SI{0.38}{\micro\meter\per\micro\meter})\cite{Bamber1981}.
\begin{equation}
\begin{aligned}
E_{mises} = \sqrt{\frac23 \left[E_{rr}^2+2+\left(-\frac12 E_{rr}\right)^2\right]} = |E_{rr}|.
\end{aligned}
\label{Emis}
\end{equation}

\subsection{Experiments in \textit{ex vivo} Tissue}

\subsubsection{Experimental Arrangement}
\begin{figure}[htpb!]
\includegraphics{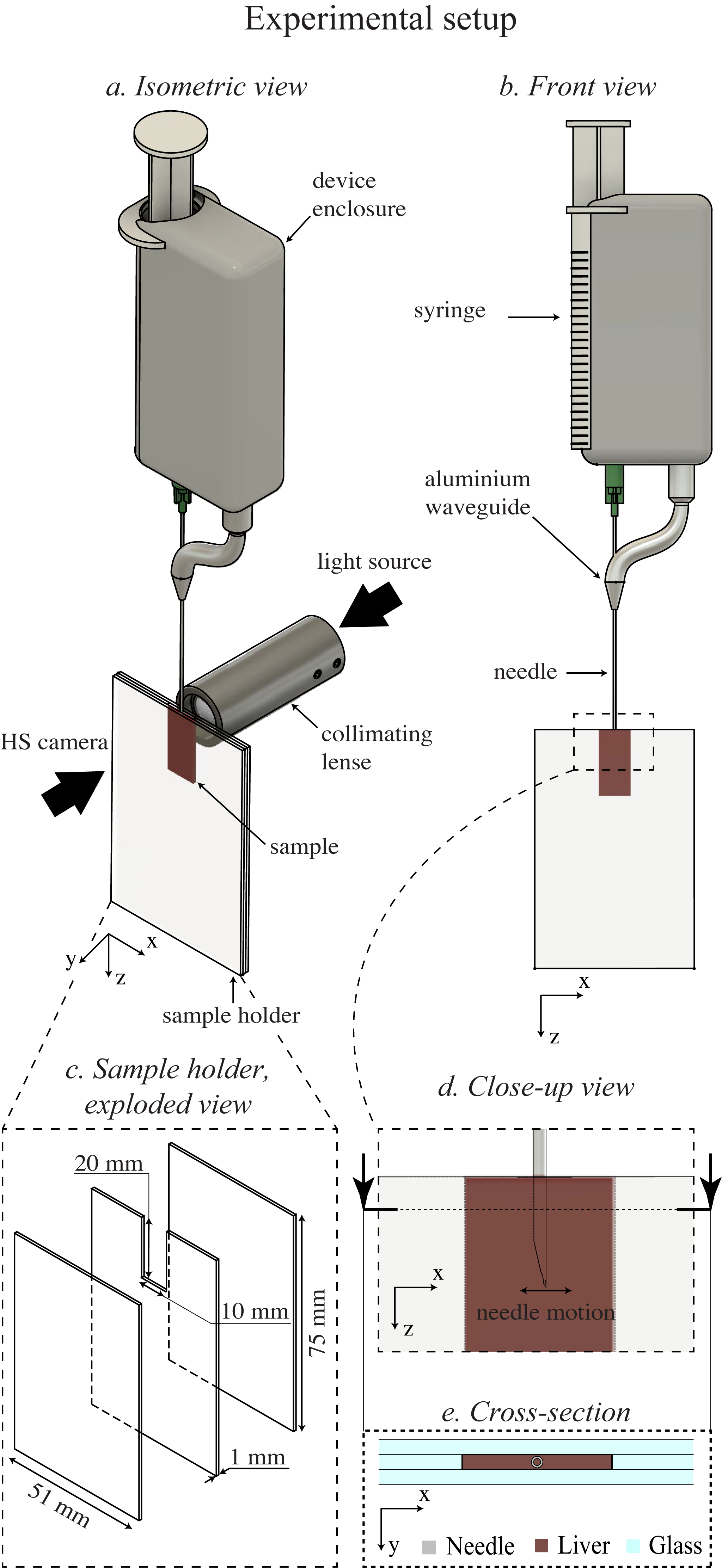}
\caption{\label{figure1} (\textbf{a},\textbf{b}) Schematics representing the experimental setup. (\textbf{c}) A custom-made glass sample holder is used as housing for a thin liver tissue slice. (\textbf{d},\textbf{e}) The ultrasonic needle is made to vibrate sideways inside the sample in the direction of the positive x-axis, while the generated cavitation events are recorded with a HS camera using a collimated beam of light to produce back-lit shadowgraph footages.}
\end{figure}

A custom-built ultrasonic device \cite{Perra2021} was used to excite a flexural vibration mode ($f = \SI{33}{\kHz}$) in a 21G hypodermic needle (length = $\SI{80}{\mm}$) (model: 4665465, 100 Sterican, B Braun, Melsungen, Germany) (Fig. \ref{figure1}a). The needle was coupled to an S-shaped 3D-printed aluminium waveguide (3D Step Oy, Ylöjärvi, Finland) that acts as a mode converter, translating the longitudinal motion provided by the Langevin transducer into a flexural motion of the needle (Fig. \ref{figure1}b). The ultrasonic device was driven by an RF amplifier (model: AG 1012LF, Amplifier/Generator, T\&C Power Conversion, Inc., Rochester, NY, United States) controlled by a function generator (model: Analog Discovery 2, Digilent, Inc., Henley Court Pullman, WA, United States). The spatial coordinates of the needle were controlled by using a motorized three-axis translation stage (model: 8MT50-100BS1-XYZ, Motorized Translation Stage, Standa, Vilnius, Lithuania). The cavitation events induced by the needle action were filmed in liver tissue using a HS camera (model: Phantom V1612, Vision Research, Wayne, NJ, United States) in combination with a macro lens (model: Canon MP-E 65 mm f / 2.8 1-5x Macro Photo, Canon Inc., Ōta, Tokyo, Japan). A collimated beam of light (model: OSL2COL, Collimation Package for OSL2IR, Thorlabs, Inc., Newton, NJ, United States), generated by a halogen fiber optic illuminator (model: OSL2IR, High-Intensity Fiber-Coupled Illuminator, Thorlabs, Inc., Newton, NJ, United States), was used to produce back-lit shadowgraph footages of the needle actuation inside tissue. 

\subsubsection{Sample Preparation}
The liver specimen (from a 26 months old female cow) was retrieved from the slaughterhouse (Vainion Teurastamo Oy, Orimattila, Finland) within 2 h \textit{post mortem} and experiments were performed within 6 h \textit{post mortem} at room temperature (22–\SI{24}{\celsius}). The specimen was first rinsed with 1×PBS (BP399-4, Phosphate Buffered Saline, 10× Solution, Fisher BioReagents, Fisher Scientific, Hampton, NH, United States) to wash away any excess of blood from its surface. Thin slices, approximately \SI{1}{mm} thick, were carefully extracted from the specimen by using a pair of microtome blades (12101840, Epredia Ultra Disposable Microtome Blades, Epredia, Portsmouth, NH, United States) fixed to a spacing of \SI{1}{mm} from each other. The liver slices were further washed in 1×PBS, cut into \SI{2}{cm}$\times$\SI{1}{cm} portions and inserted into a custom made glass sample holder (Fig. \ref{figure1}c). The sample holder was created by cutting a \SI{2}{cm}$\times$\SI{1}{cm} portion of glass from the upper part of a \SI{51}{\milli\meter}$\times$\SI{75}{\milli\meter} microscope slide (J1800BMNZ,Epredia SuperFrost Plus Adhesion slides, Special Size, Epredia, Portsmouth, NH, United States), which was placed between two intact microscope slides, in order to form a pocket for the tissue sample (Fig. \ref{figure1}c,d). 

\subsubsection{Data Acquisition}
Since the penetration depth $\delta$ of light into bovine liver is estimated to be, for example, \SI{1.44}{\mm} for a wavelength of \SI{635}{\nm}\cite{Arslan2018}, the thickness of the sample and the light source spectrum were considered appropriate to ensure a good visibility of the needle inside tissue during the HS recordings. During the experiments, the needle was first carefully inserted into the specimen at a depth of \SI{5}{\mm} and penetration speed of \SI{50}{\micro\meter\per\second}. Ultrasound waves (\SI{33}{\kHz}, pulse repetition frequency (PRF) = \SI{55}{\Hz}, duty cycle (DC) = \SI{50}{\%}) at 3 different TAP levels (0.2 W (\textit{n} = 5), 0.5 W (\textit{n} = 5) and 0.8 W (\textit{n} = 5)) were then applied to the device, while the needle movement inside tissue was recorded with the following settings: sample rate = \SI{130000}{fps}, exposure = \SI{7.1}{\micro\second}, resolution = 256 pixels × 256 pixels, lens aperture = 2.8, spatial resolution = \SI{5.5}{\micro\meter\per pixel}.

\subsubsection{Data Analysis}
The HS frames were analysed in MATLAB (R2020b) \cite{Mathworks2016} to quantify the projected area of cavitation and the needle displacement using a similar method presented in our recent publication\cite{Perra2021}. A cross-correlation based image registration was performed along the x-axis between the reference frame $I_1$ and the the i-th frame $I_i$, in order to estimate the needle displacement $\Delta x_i$ from its reference position. The image $I_1$ was then rigidly translated by $\Delta x_i$ and thresholded with the Otsu method\cite{Otsu1996}, while an Otsu thresholding followed by a morphological closing operation (circular structuring element, diameter = 7 pixels) was applied to the image $I_i$. The segmented image $I_{cav,i}$ showing only the cavitation activity is obtained by subtracting the binarized reference image $I_{bw,1}$ from the closed image $I_{bw,i}$. Since the needle shape in $I_{bw,1}$ does not perfectly match the one in $I_{bw,i}$, a final morphological opening operation (circular structuring element, diameter = 3 pixels) was applied to the output $I_{cav,i}$ in order remove any pixels that may have remained after the subtraction operation and that are not representative of the cavitation activity. 
 
 Probability maps (Eq. (\ref{probmap})), showing the probability of cavitation manifesting around the needle tip, and projected areas of cavitation activity over time (Eq. (\ref{cavarea})) were finally calculated as follows:
 \begin{equation}
\begin{aligned}
P_{cav} = \frac{100}{N}\sum_{i=1}^N{I_{cav,i}},
\end{aligned}
\label{probmap}
\end{equation}
\begin{equation}
\begin{aligned}
A_{cav,i} = \int\int{I_{cav,i}dx dy},
\end{aligned}
\label{cavarea}
\end{equation}
where $N = 25000$ is the total number of frames. 

Velocity maps, shear and strain rate maps were generated using the PIVlab toolbox \cite{Thielicke2021}.

\section{Results}

\subsection{Simulation of Cavitation in Liver}
Based on the simulations, the ultrasonic action of the needle induced expansion of bubbles in the proximity of the needle tip, when the needle was driven by ultrasonic waves at the frequency of 33 kHz (Fig. \ref{figure3}a.1). The TAP levels were 0.2, 0.5 and \SI{0.8}{\W}, which helped generate driving pressure waves with peak amplitudes of 90, 500 and \SI{900}{\kPa} evaluated at the location \textbf{B1} (Fig. \ref{figure3}a.2). The cavitation bubble dynamics evaluated at location \textbf{B1} exhibited a TAP dependent behaviour. At the lowest TAP level, the bubble oscillated around its initial radius $R_0$ with velocities lower than \SI{1}{\meter\per\second}. At increased TAP levels, the bubble radius expanded up to 45 and \SI{100}{\micro\meter}, reaching a maximum velocity of 30 and \SI{100}{\meter\per\second} during the collapse (Fig. \ref{figure3}a.2) at employed TAP levels of 0.5 and \SI{0.8}{\W}, respectively. 

The probability maps presented in Fig. \ref{figure3}b.1 show the chance for cavitation events to occur around the needle tip, defined as when the radius $R$ of a cavitation nucleus becomes as large as $2R_0$\cite{Gaitan2005}. According to this criterion, no cavitation activity was detected at \SI{0.2}{\W}, while the probability of cavitation occurrence became higher at increased TAP levels, being up to 50\% at \SI{0.8}{\W}. 

The predicted volume of tissue influenced by cavitation activity, calculated by identifying the regions where $E_{mis} > 0.38$, was almost zero at low TAP, suggesting that no important deformations were induced in the surrounding tissue (Fig. \ref{figure3}b.2). However, higher strains can be generated in tissue at higher TAP levels due to the elevated cavitation bubble activity, leading to an increase in the influenced volume of tissue by cavitation up to \SI{3.2}{\cubic\milli\meter} after 2 acoustic cycles at \SI{0.8}{\W}.

\subsection{Observation of Cavitation Events in \textit{ex vivo} Liver}
Thin slices of liver tissue were sonicated at different TAP levels (0.2 W (\textit{n} = 5), 0.5 W (\textit{n} = 5) and 0.8 W (\textit{n} = 5)) with the ultrasonically actuated needle. Fig. \ref{figure4}a shows some exemplary frames acquired with the HS camera, when a halogen fiber optic light source was used to produce shadowgraph images of the needle movement inside tissue. When the delivered TAP was \SI{0.2}{\W}, no cavitation activity was detected. At the TAP level of \SI{0.5}{\W}, the needle motion induced the formation of cavitation bubbles, which mostly took place at the distal end of the needle. However, when the delivered TAP was increased to \SI{0.8}{\W}, multiple cavitation bubbles can be noticed along the needle tip, extending to a few hundreds of \SI{}{\micro\meter} from the needle boundaries along the directions parallel to the needle motion. Fig. \ref{figure4}b shows the cavitation probability maps calculated across the entire duration of the HS footages, which suggest that no cavitation events were observed at the lowest TAP \SI{0.2}{\W} employed. By increasing the TAP, the probability of seeing cavitation bubbles was up to 10\% in the region within \SI{100}{\micro\meter} from the needle tip along the positive x-axis and \SI{300}{\micro\meter} along the negative z-axis, while this region became considerably greater in area and uniformly distributed around the needle tip, when the highest TAP of \SI{0.8}{\W} was employed. 

Fig. \ref{figure5}b represents the time evolution of the needle tip peak displacement, obtained by computing the moving maximum of the raw data and using a window with a size of approximately 2 acoustic cycles (\SI{60}{\micro\second}). In all experiments, the peak displacement reached its maximum value within the first burst, being $\sim$ 9, 45 and \SI{100}{\micro\meter} at the TAP levels of 0.2, 0.5 and \SI{0.8}{\W}, respectively. Fig. \ref{figure5}c shows the projected area of the cavitation activity (filtered with a moving average, window size $\sim$10 acoustic cycles) as a function of time. It can be noted that no cavitation activity was present at \SI{0.2}{\W}, while some activity was detected at \SI{0.5}{\W}, and, at \SI{0.8}{\W}, the measured cavitation activity was relatively elevated. 

Measurements of needle displacement and cavitation activity exhibited high repeatability within the same power groups, as shown in Fig. \ref{figure5}d. In the first burst, the needle tip peak displacements were \SI[multi-part-units = single]{9.5 (64)}{\micro\meter} (average $\pm$ standard deviation, \textit{n} = 5), \SI[multi-part-units = single]{34.3 (10)}{\micro\meter} and \SI[multi-part-units = single]{61.6 (19)}{\micro\meter}, when the employed TAP levels were 0.2, 0.5 and \SI{0.8}{W}, respectively. The needle tip peak displacement stabilised within 3 bursts, reaching the values of \SI[multi-part-units = single]{10.8 (85)}{\micro\meter} (\SI{0.2}{W}), \SI[multi-part-units = single]{41.5 (14)}{\micro\meter} (\SI{0.5}{W}) and \SI[multi-part-units = single]{91.9 (23)}{\micro\meter} (\SI{0.8}{W}) after 10 bursts. Fig. \ref{figure5}e shows the time integral of cavitation activity calculated for each individual burst. In all experiments, no cavitation activity was recorded for TAP = \SI{0.2}{\W}, while it slowly built-up over time at TAP = \SI{0.5}{\W}, being \SI[multi-part-units = single]{0.014 (2)}{\square\milli\meter\milli\second} after the first burst, and reaching the value of \SI[multi-part-units = single]{0.055 (12)}{\square\milli\meter\milli\second} during the 10th burst. At the highest TAP employed, the cavitation level observed during the first burst was \SI[multi-part-units = single]{0.177 (60)}{\square\milli\meter\milli\second}, and reached its maximum intensity in the last burst (\SI[multi-part-units = single]{0.386 (16)}{\square\milli\meter\milli\second}). Overall, the temporally local peak displacement of the needle tip measured for each TAP level were \SI[multi-part-units = single]{11.3 (68)}{\micro\meter}, \SI[multi-part-units = single]{44.4 (15)}{\micro\meter} and \SI[multi-part-units = single]{97.2 (18)}{\micro\meter}, which led to total cavitation activity values of $\sim$ \SI[multi-part-units = single]{0 (0)}{\square\milli\meter\milli\second}, \SI[multi-part-units = single]{0.459 (32)}{\square\milli\meter\milli\second} and \SI[multi-part-units = single]{3.44 (78)}{\square\milli\meter\milli\second} for TAP = 0.2, 0.5, \SI{0.8}{\W}, respectively.

Velocity maps were generated out of 2 consecutive frames of the HS videos in order to estimate the velocity field of the tissue at the moment of a cavitation bubble collapse. Fig. \ref{figure6}a shows the velocity vector field distribution overlapped to a HS frame showing a cavitation event, when the highest TAP is employed. The velocities are the highest at the very tip of the needle, being approximately \SI{3}{\meter\per\second} in this region (Fig. \ref{figure6}b). Importantly, according to the simulation, the velocity of the tissue-air -interface can be remarkably greater, \textit{i.e.} up to \SI{100}{\meter\per\second}. However, the limited frame rate adopted during the recordings (130000 fps) did not permit capture of the very moment of the cavitation collapse, which resulted in underestimation of its maximum velocity. Fig. \ref{figure6}c shows the shear rate distribution around the needle, being the highest in magnitude (\SI{20}{\per\milli\second}) at the proximity of the cavitation bubble boundary. This is reasonable since the cavitation bubble deformation is known to exert considerably high shears and stresses in the surrounding medium. In Fig. \ref{figure6}d the strain rates assume negative values, which denote a compression state, on the left hand side of the needle and positive values in the proximity of the cavitation bubble. The needle movement is in the direction of the negative x-axis, which causes the adjacent portion of tissue to be compressed on the left-hand side of the needle, and to be stretched on the right-hand side. 

\section{Discussion}
The results indicate that cavitation events can be triggered by actuating a standard medical needle with ultrasonic flexural waves in liver tissue at the frequency of \SI{33}{\kHz}. The numerical results suggested that the cavitation activity mostly took place at the needle tip, which was optically confirmed with HS photography. This is explained by the flexural vibration mode induced in the needle, which makes the needle oscillate with its highest displacements at its tip, thus enabling higher pressure amplitudes in this region. Since cavitation is a strictly related threshold phenomenon, cavitation events are most likely to appear at the needle tip location, where most of the acoustic intensity is concentrated. Moreover, due to the geometric spreading of the acoustic wavefront, directed outwards from the needle shaft, the acoustic intensity decays rapidly further away from the needle, hence limiting the cavitation effects to the proximity of the needle tip. However, by looking at a cross section of the numerical model, we noticed that a greater number of cavitation nuclei seemed to interact with the acoustic field within the needle cannula. This has to do with the concave shape of the inner walls, which tends to focus the acoustic energy towards the center axis of the needle. Also, the acoustic wave is reflected within the inner walls of the needle, which might contribute to raising the magnitude of the acoustic intensity in this region. This justifies the elevated cavitation activity in this region, which causes the portion of tissue located at the needle opening to be influenced the most (Fig. \ref{figure3}b.2). However, the volume of tissue influenced by the cavitation activity seems to decrease drastically at lower TAP levels. 

The experimental results showed that the probability of triggering cavitation events in soft tissue is a function of TAP, suggesting the existence of a threshold ($\SI{0.2}{\W}<$ TAP $<\SI{0.5}{\W}$ ) for enabling cavitation, when a standard medical needle is actuated in soft tissue. Even though the cavitation threshold value was not confirmed through a direct measurement, based on the simulation results, this threshold seems to be approximately \SI{-500}{\kPa}. Assuming an initial size of the cavitation nuclei of \SI{500}{\nm}, the natural frequency of a bubble with this radius would fall in the MHz range\cite{Atchley1989}.
Since the excitation frequency $f_0$ = \SI{33}{kHz} used in this study is far below the resonance frequency $f_n$ of the bubble, the cavitation threshold criterion is governed by the Blake pressure\cite{Atchley1989}, which determines the critical negative pressure below which a cavitation event will occur:
\begin{equation} 
\begin{aligned}
P_B = P_\infty + \frac{8\sigma}{9}\sqrt{\frac{3\sigma}{2R_B^3(P_\infty+(2\sigma/R_B))}},
\end{aligned}
\label{blake}
\end{equation}
where $P_B$ is the Blake pressure, $\sigma$ is the surface tension and $R_B$ is the Blake bubble radius. Under these assumptions, the Blake threshold for a bubble of $R_0 = \SI{500}{\nm}$ is $\sim$ \SI{-200}{\kPa}, which is in line with the numerical results, in which the estimated threshold was around \SI{-500}{\kPa}.

The interpretation of the results here presented is of fundamental importance in the context of different medical applications. In our recent study, the influence of the ultrasonic action of a medical needle was exemplified in liver tissue by comparing the yield mass collected with the USeFNAB technique to the one obtained with the conventional FNAB approach. The major finding was that, by increasing the TAP level, the yield of a liver biopsy was increased up to 5$\times$as compared to when a standard FNA was performed, without inducing major alterations to the sample quality up to TAP of 0.8 W. More importantly, a TAP of $\SI{0.2}{W}$ was enough to increase the biopsy yield by almost 2$\times$. Based on the findings of the present study (Fig. \ref{figure4} and Fig. \ref{figure5}), it seems that this TAP level is unlikely to generate cavitation events in liver, indicating that the tissue yield increase observed at this TAP can be in part associated with the tissue cutting mechanisms arising from shear and hydrodynamic effects promoted by the ultrasonic vibration of the needle tip, rather than being induced by cavitation. However, higher TAP levels allowed us to obtain even larger tissue sample masses\cite{Perra2021} as well as more frequent cavitation activity. Although a clear correlation between the cavitation activity and the tissue yield is yet to be proven, these observations do not exclude the possibility for cavitation to be contributing to the enhancement of tissue collection. In fact, the high strain rates generated in the proximity the gas/tissue interface can potentially induce different viscoelastic mechanical responses (namely stiffening, softening, hardening and tissue failure observed in porcine liver under high strain rate compression testings\cite{Chen2009a}) that might facilitate the tissue cutting mechanisms that yield an increase in sample extraction. 

Regarding the safety aspects of the biopsy application in relation to the potential cavitation-induced effects in tissue, one should consider the mechanical index MI = $P_r/\sqrt{f_c}$, where $P_r$ is the peak negative pressure (MPa) and $f_c$ is the excitation frequency (MHz). According to our simulations, at the lowest TAP employed MI $\sim$ 0.4, which would ensure a cavitation-free biopsy procedure, since cavitation is unlikely to take place at MI $<$ 0.5\cite{Apfel1991}. At \SI{0.5}{\W} the MI is approximately 2.7, which will most likely induce the formation of cavitation bubbles; this might impact on the safety, as at MI values greater than 1.9 potential bioeffects might be induced in the tissue\cite{Sen2015}. These bioeffects may include cell lysis and extravasation of blood\cite{Hallow2006a,Church2008a,Haar2010}. The highest TAP employed should be avoided for biopsy applications, as the high MI value (4.9) suggests that bioeffects and tissue damage due to the bubble collapses are likely to appear.

If uncontrolled, cavitation events can lead to deleterious effects in soft tissue. However, this could be turned into a therapeutic advantage if one aims to treat unhealthy tissue, such as tumors. At high levels (TAP $>$ \SI{0.8}{\W}), the needle vibration is anticipated to cause the formation of large clouds of cavitation bubbles and elevated tissue heating, which may arise from the viscous friction forces that can appear at the bubble surface. These effects can be used in medical applications such as tumor ablation \cite{Zhou2011}, histotripsy or lithotripsy \cite{Yoshizawa2009}, where the medical intent is to achieve a complete or partial destruction of the target by mechanical and thermal means. Since a fine hypodermic needle is employed to bring the acoustic energy directly into the target, one may be able to easily access different locations inside the body to provide minimally invasive treatment of solid organ cancers of the prostate \cite{Chen2016}, thyroid \cite{Pacella2013} and pancreas \cite{Larghi2021}, and other lesions too\cite{Piccioni2019}. In addition, the cavitation phenomena generated with this technology could potentially find use in other applications as a way to ultrasonically activate sonosensitive carriers for the release of drugs\cite{George2019,Husseini2005,Zhou2014a}, mediate drug or gene delivery into cells\cite{Dachs1997}, or to improve the permeation of tissue allowing the entry of therapeutic agents (\textit{e.g}. as with ultrasonically mediated blood-brain barrier opening)\cite{Aryal2014}. 

The limitations of this study include the inability to replicate a numerical model representative of the real-world scenario. The equations adopted in the simulations are highly parameter-dependent, and since some of the viscoelastic and acoustic properties of tissue are largely unknown, some assumptions had to be made, for example, on the tissue viscosity, surface tension and initial radius of cavitation nuclei. For simplicity, the interactions between individual bubbles were neglected as well as potential temperature related effects, which, conversely, are known to play an important role in the context of ultrasonic cavitation in soft tissue. Moreover, since the use of thin portions of liver tissue were necessary to visualize the needle and cavitation activity during the experiments, the setup might not replicate the same acoustic and mechanical conditions as in an FNAB procedure. 

Nevertheless, the presented results offer an understanding of the cavitation phenomena in liver tissue near the ultrasonically actuated medical needle. Such findings could serve as a starting point for designing and developing an ultrasonic biopsy device in compliance with the safety standards for clinical applications, and for exploring its potential in other medical applications involving pathological destruction of tissue.

\section{Conclusion}
To conclude, we have studied numerically the dynamics of cavitation bubbles generated in liver tissue near the tip of an ultrasonically actuated needle. Experimentally, we have developed a method to capture and quantify the cavitation activity within thin slices of fresh bovine liver. The main finding was that cavitation exhibited a TAP dependent behavior, manifesting at TAP $>$ \SI{0.2}{W} and with intensity proportional to the TAP level. Based on a qualitative comparison, the numerical and the experimental results presented similarities concerning the cavitation threshold and the spatial probability of cavitation occurrence around the needle tip. The results are important since they broaden the understanding of the onset and spatio-temporal behavior of cavitation near ultrasonically actuated medical needles. This is especially relevant for ensuring appropriate safety in clinical scenarios, but also for employing the information in the development of USeFNAB and other new applications of ultrasonically actuated medical needles. 

\begin{acknowledgments}
We thank all members of the Medical Ultrasonics Laboratory (MEDUSA) at Aalto University and Prof. Edward Hæggström at University of Helsinki for constructive discussions related to the topic. The Academy of Finland is acknowledged for financial support (grants 314286, 311586 and 335799). 
\end{acknowledgments}

\section*{Author Contributions Statement}
All authors contributed to the design of the study, writing or reviewing the manuscript and have approved the final version of the manuscript. Emanuele Perra produced all data and conducted the data analyses.

\section*{Conflict of Interest}
Heikki J. Nieminen, Kenneth P.H. Pritzker have stock ownership in Swan Cytologics Inc., Toronto, ON, Canada and are inventors within the patent application WO2018000102A1. Emanuele Perra and Nick Hayward do not have any competing interests in relation to this work. 

\section*{Data Availability}
The datasets are available upon request.

\section*{Code Availability}
The codes are available upon request.

\bibliography{references}
\newpage

\begin{table}[]
\caption{A table of the general parameters of the numerical model at the ambient temperature of \SI{25}{\degreeCelsius}.}
\label{Table1}
\begin{tabular}{@{}lll@{}}
\toprule
Properties                            & Values                         & References                \\ \midrule
Ultrasound frequency, $f$             & \SI{33}{\kHz}                  & -                         \\
Speed of sound, \textit{$c_{\infty}$} & \SI{1575}{\meter\per\second}   & \cite{Amin1989}           \\
Density, $\rho_{\infty}$              & \SI{1060}{\kg\per\cubic\m}     & \cite{Pahk2019}           \\
Nonlinearity parameter, $B/A$         & 7.14                           & \cite{Bamber1981}         \\
Attenuation coefficient, $\alpha$     & \SI{1.39}{\deci\bel\per\meter} & \cite{Amin1989}           \\
Ambient pressure, $p_{\infty}$         & \SI{101.325}{\kPa}             & \cite{Mancia2019}         \\
Polytropic index, $\kappa$            & 1                            & \cite{Mancia2017}         \\
Surface tension, $S$                  & \SI{56}{\milli\N}              & \cite{Warnez2017}         \\
Viscosity, $\mu$                      & \SI{30}{\milli\Pa\s}           & \cite{Warnez2017}         \\
Shear modulus, $G$                    & \SI{1.8}{\kPa}                 & \cite{Palmeri2008}        \\
Initial bubble radius, $R_0$          & \SI{500}{\nm}                  & \cite{Vlaisavljevich2015} \\ \bottomrule
\end{tabular}%
\end{table}

\begin{figure*}[htpb!]
\includegraphics[width=\textwidth]{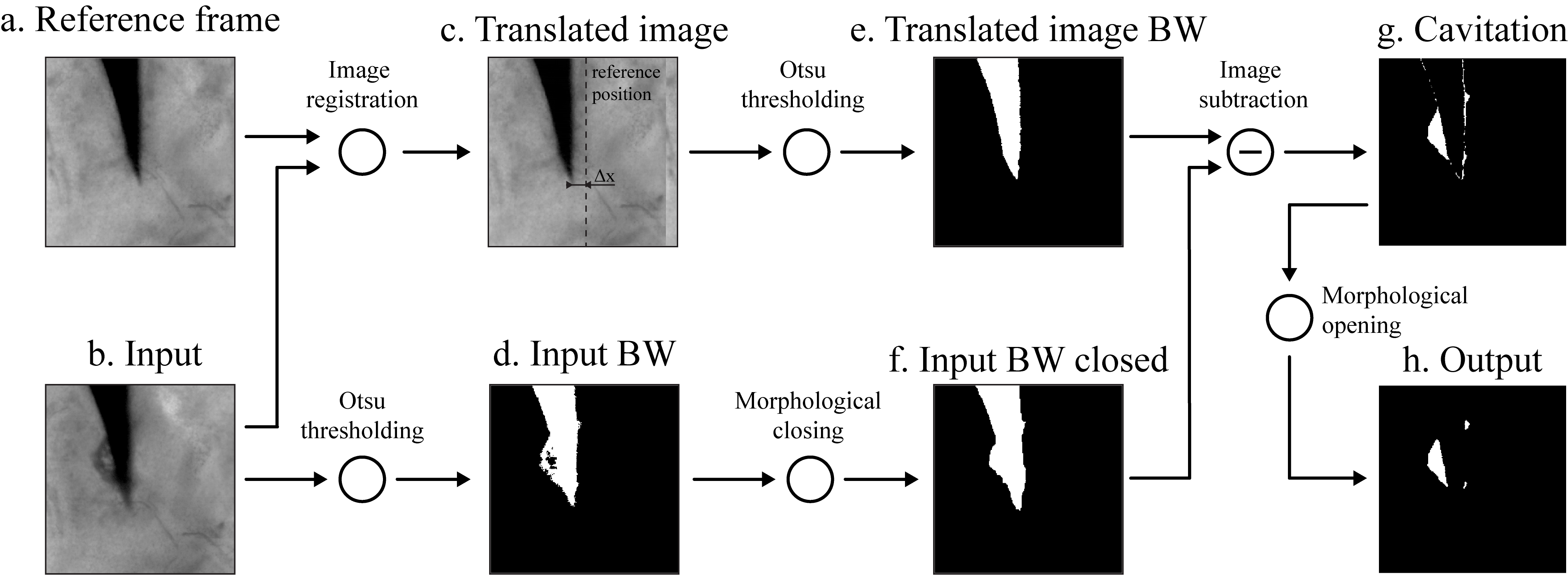}
\caption{\label{figure2} General description of the segmentation process of the HS video frames. A cross-correlation based image registration is applied along the x-axis between (\textbf{a}) the reference frame and the (\textbf{b}) i-th frame, in order to estimate the displacement $\Delta x_i$ of the needle tip from its reference position. (\textbf{c}) The reference frame is then translated horizontally by $\Delta x_i$, while (\textbf{d}) the input frame is thresholded with the Otsu method. (\textbf{e}) The binary image showing the translated reference frame is subtracted to (\textbf{f}) the thresholded input frame, which was previously closed with a morphological closing operation. (\textbf{g}) The result of the subtraction is finally filtered with a morphological opening operation, in order to produce (\textbf{h}) the binary mask for the cavitation activity.}
\end{figure*}

\begin{figure*} [htpb!]
\includegraphics[width=\textwidth]{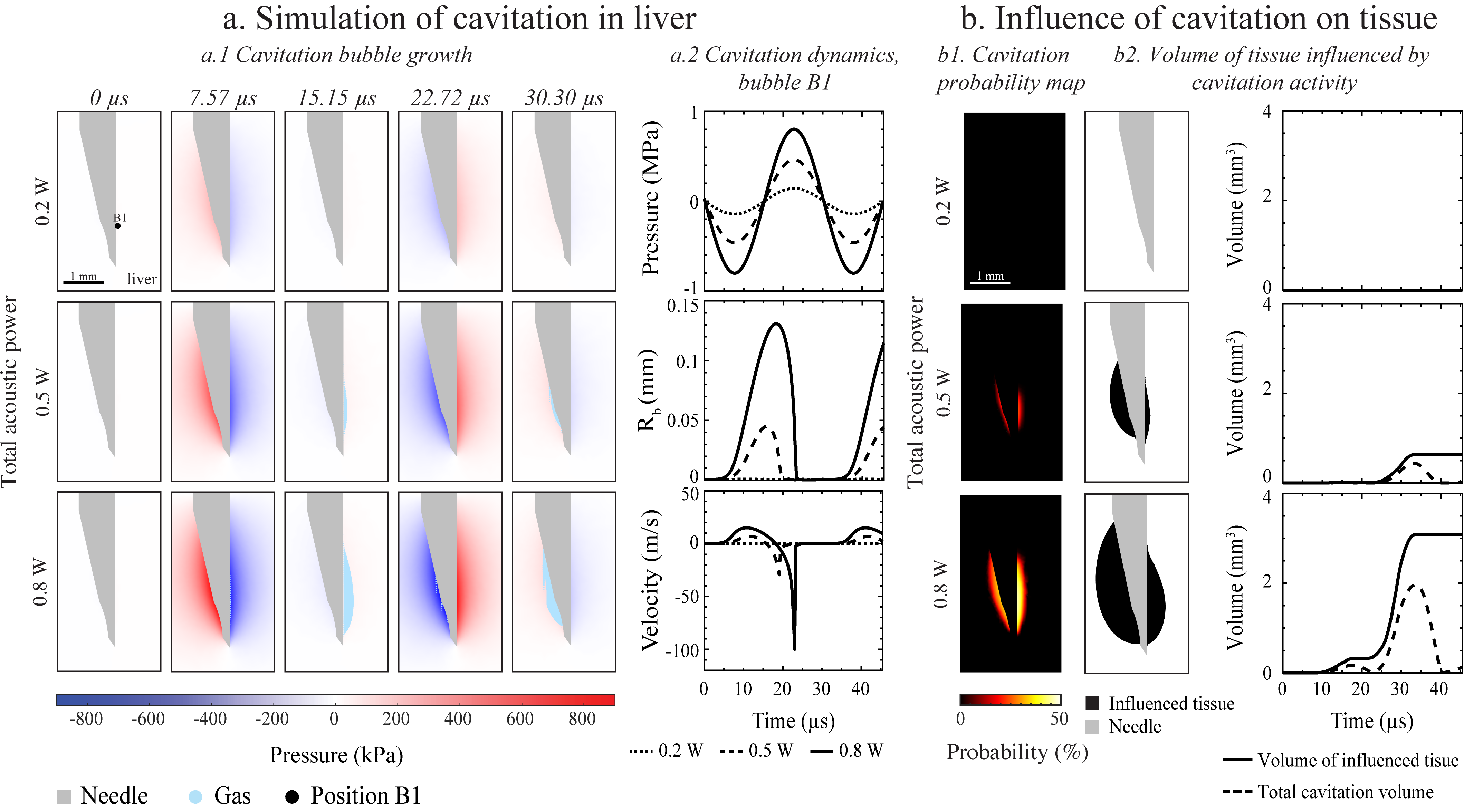}
\caption{\label{figure3}(\textbf{a}1) Simulation of cavitation around the tip of an ultrasonic hypodermic needle embedded in liver tissue, where the selected time points represent fractions of an acoustic cycle of duration $T=1/f$. The surrounding medium is seeded with cavitation nuclei and the needle is actuated at TAP levels of 0.2, 0.3 and \SI{0.8}{\W} at a frequency of \SI{33}{\kHz}. (\textbf{a}2) Evolution of the driving pressure (MPa), radius (mm) and velocity (m/s) of a bubble evaluated at position \textbf{B1}. (\textbf{b}1) The probability maps show the cavitation occurrence probability around the needle tip, when TAP levels of 0.2, 0.3 and \SI{0.8}{\W} are delivered to the needle. (\textbf{b}2) The amount of tissue volume influenced by cavitation activity is estimated by identifying the regions where the von Mises strain exceeds the ultimate fractional strain measured for liver (0.38 $\upmu$m/$\upmu$m). The numerical results suggest that measurable cavitation events are triggered at TAP $>$ \SI{0.2}{\W}, with greater influence on tissue at higher TAP levels than at lower TAP levels.}
\end{figure*}

\begin{figure} 
\includegraphics{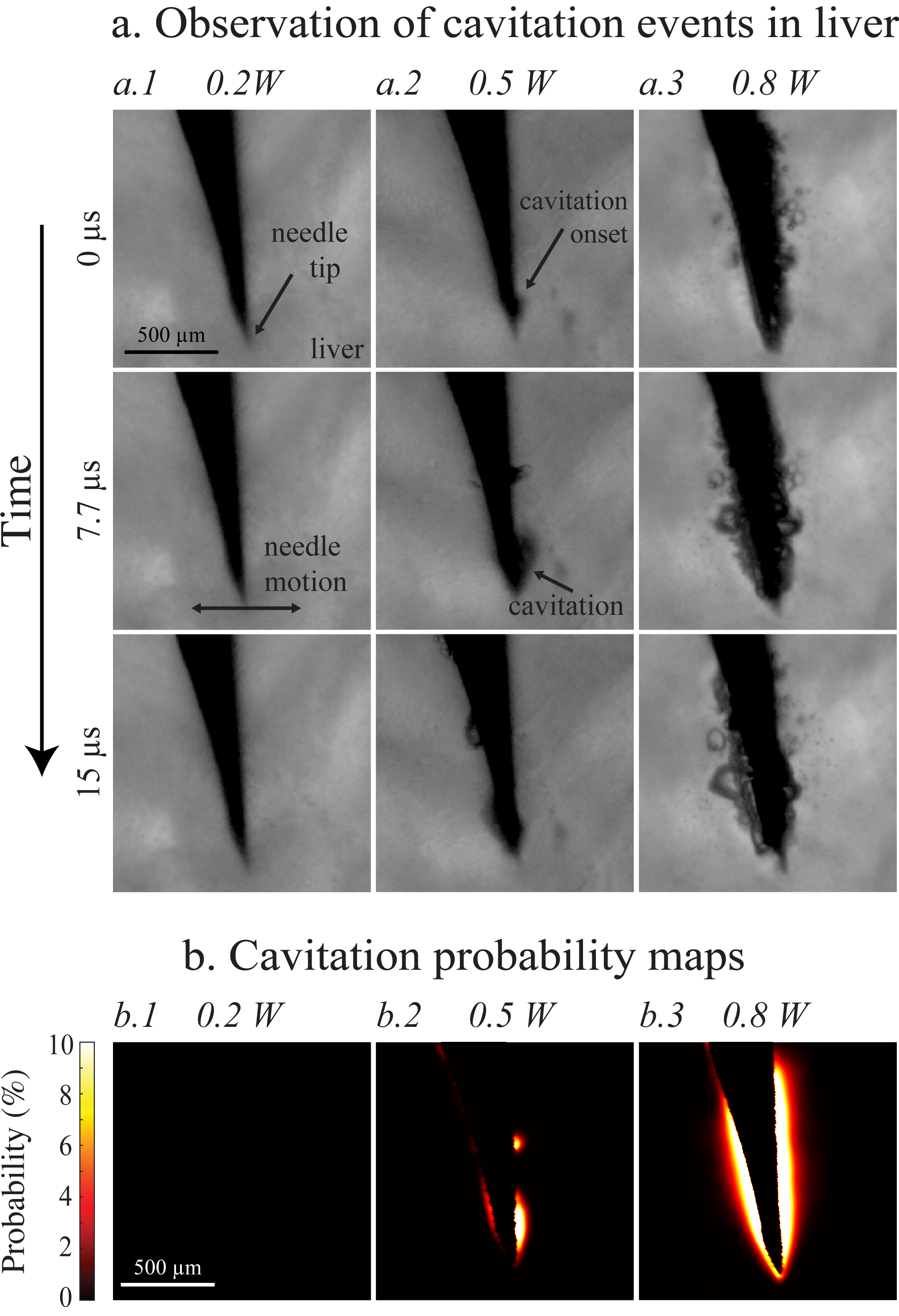}
\caption{\label{figure4}(\textbf{a}) Exemplary HS video frames showing cavitation events taking place at the needle tip at 3 different TAP levels. (\textbf{b}) The probability maps demonstrate that no cavitation activity is recorded at \SI{0.2}{\W}, while it is more frequent at higher TAP levels.}
\end{figure}

\begin{figure*}[htpb!]
\includegraphics[width=\textwidth]{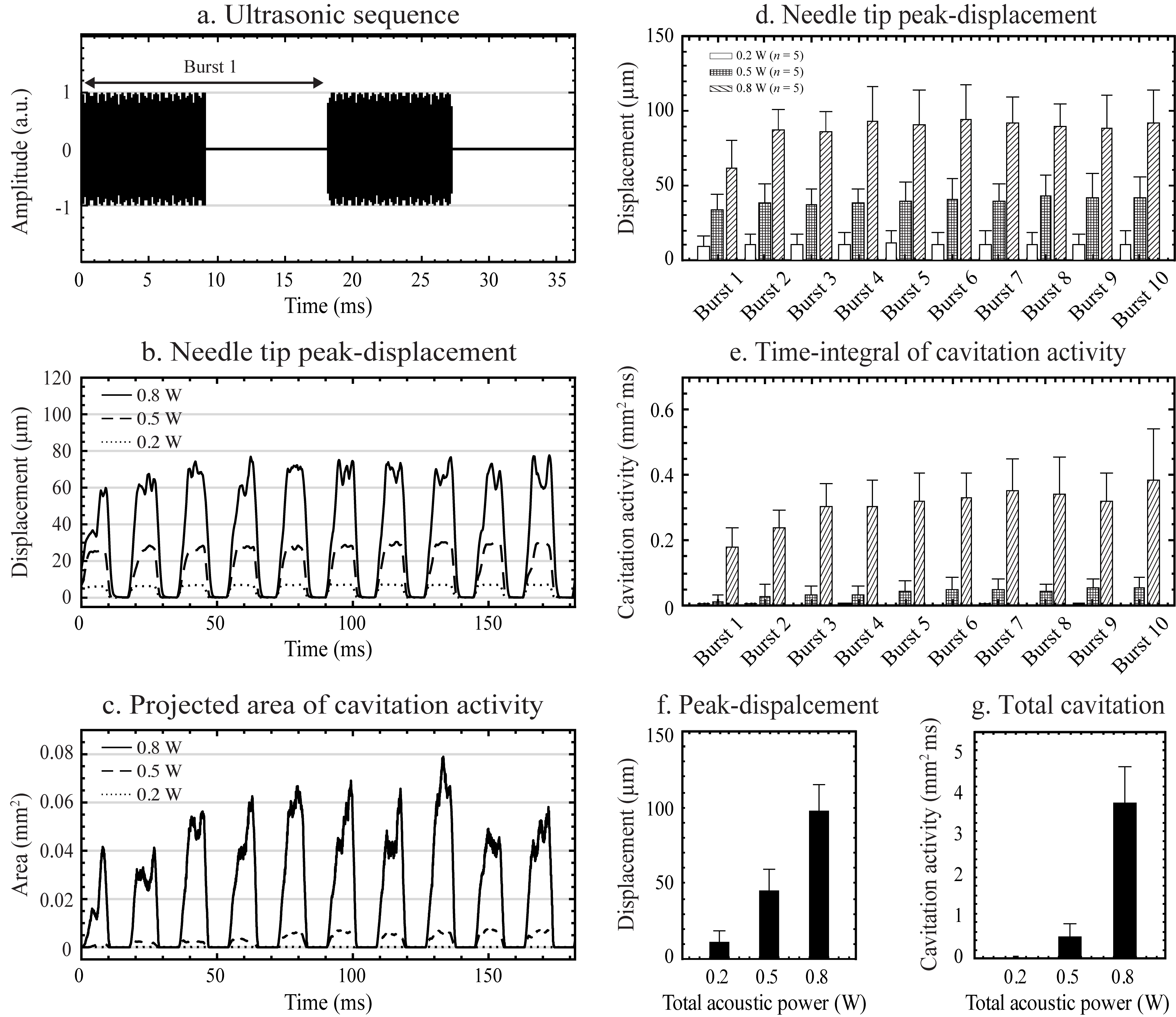}
\caption{\label{figure5} (\textbf{a}) Ultrasonic sequence adopted during the experiments ($f$ = \SI{33}{\kHz}, pulse repetition frequency (PRF) = \SI{55}{\Hz}, duty cycle (DC) = \SI{50}{\%}). (\textbf{b}) Time evolution of the needle-tip displacements, obtained by computing the moving maximum of the raw data (window size $\sim$ 2 acoustic cycles), and (\textbf{c}) projected area of the cavitation activity (filtered with moving average, window size $\sim$10 acoustic cycles). \textbf{d} and \textbf{e} represent the peak-tip displacement and time integral of cavitation activity calculated for each individual burst of the ultrasonic sequence, while \textbf{f} represents the same information evaluated across a time window of \SI{180}{\ms}. In the bar charts, the bar height represents the mean of the data set and the error bar indicates the standard deviation.}
\end{figure*}

\begin{figure}[htpb!]
\includegraphics{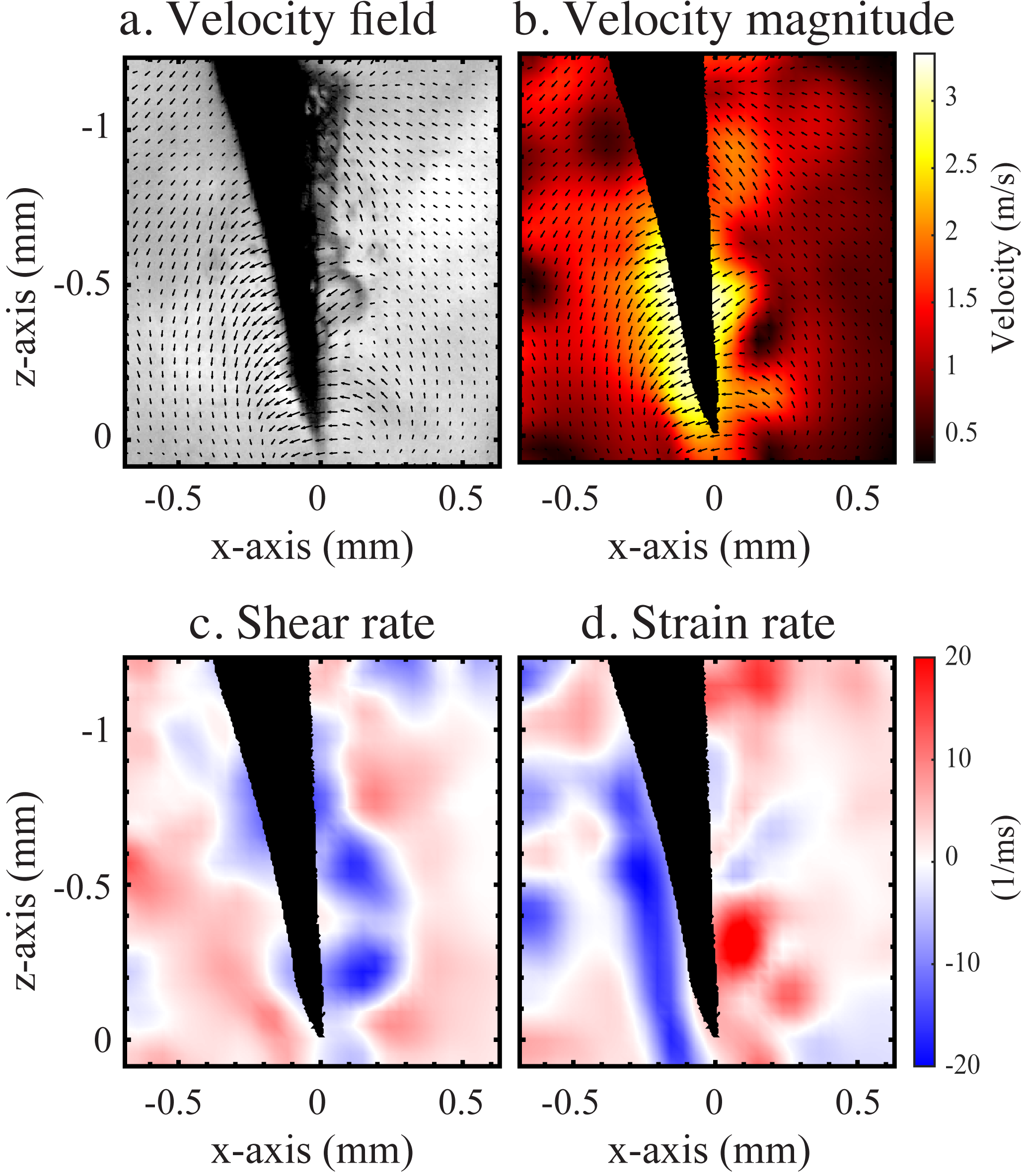}
\caption{\label{figure6} (\textbf{a}) Velocity vector field distribution overlapped to an exemplary HS frame showing a cavitation event and (\textbf{b}) the velocity magnitude map, when the highest TAP (\SI{0.8}{\W}) is employed. \textbf{c} and \textbf{d} show the shear and strain rate distribution in the tissue surrounding the needle tip.}
\end{figure}

\end{document}